\documentclass[utf8]{frontiersFPHY} %
\usepackage{url,lineno,microtype,subcaption}
\usepackage[onehalfspacing]{setspace}

\def\keyFont{\fontsize{8}{11}\helveticabold }
\def\firstAuthorLast{J. Rotureau} 
\def\Authors{J. Rotureau}
\def\Address{NSCL/FRIB Laboratory, Michigan State University, East Lansing, Michigan 48824, USA}
\def\corrAuthor{J. Rotureau}
\def\corrEmail{jjmrotureau@gmail.com}

\begin{document}
\onecolumn
\firstpage{1}

\title[Coupled-cluster computations of optical potential for medium-mass nuclei]{Coupled-cluster computations of optical potential for medium-mass nuclei}

\author[\firstAuthorLast ]{\Authors} 
\address{} 
\correspondance{} 

\extraAuth{}

\maketitle

\begin{abstract}
  Recent progress in the numerical solution of the nuclear many-body problem
  and in the  development of nuclear Hamiltonians rooted in Quantum Chromodynamics, has opened the door
  to first-principle computations of nuclear reactions.  
  In this article, we discuss the current status of {\it ab initio} calculations of nucleon-nucleus optical potentials
  for medium-mass systems, with a focus on results obtained with the coupled-cluster method. 
%
%
\tiny
\keyFont{ \section{Keywords:} nuclear reactions, nuclear structure, optical potential, ab-initio method, Green's function, chiral
  effective field theory.}
\end{abstract}
\section{Introduction}
Understanding the structure and dynamics of atomic nuclei in terms of nucleons and their mutual
interactions is one of the main goals of nuclear physics. At the typical energy scale of nuclear phenomena,
the quarks and gluons degrees of freedom are not resolved. As a consequence, in  this context, nucleons can be treated as point-like particles
and the nuclear problem with protons and neutrons can be viewed as a low-energy effective approximation to QCD.
Within the framework of Effective Field Theory (EFT), inter-nucleon interactions consistent with the chiral symmetry
can nowadays be derived systematically in terms of nucleon-nucleon, three-nucleon, and higher many-nucleon forces  \cite{beane,epel,epel2,mach,n2losat,nog}. Starting with a given Hamiltonian, {\it ab initio} calculations of nuclei aim at  solving the many-body Schr\"odinger equation
without any uncontrolled approximations. Within the last decades, the increase in computing power and
the development of powerful many-body methods, combined with the use of chiral-EFT interactions, have enabled a quantitative description of light and medium-mass nuclei 
{\it ab initio} \cite{no_core0,qmc0,scgf0,cc0,srg0,lattice0}.
With the inclusion of continuum effects in many-body methods, {\it ab-initio}
calculations have also reached parts of the nuclear chart far from stability where the coupling to continuum states
and decay channels plays an important part in the structure of nuclei \cite{hagen_cc_gam,papadimitriou2013,hagen_oxy,CC_17F,cc_ni78,nav_7he,nav_9be,nav_9He,srg_gam}).

A lot of progress has been made as well in the development of {\it ab initio} methods for nuclear reactions.
The No-Core Shell Model with the Resonating Group Method (NCSM/RGM) or with continuum (NCSMC) have successfully
described scattering and transfer reactions for light targets \cite{rgm1,rgm2,rgm3},
the Green's Function Monte Carlo \cite{gfmc_scat1,gfmc_scat2} has recently been applied to nucleon-alpha scattering using
chiral NN, 3N forces \cite{gfmc_scat3}, and  lattice-EFT computations of alpha-alpha scattering have recently
been reported \cite{elhatisari2015}.
For medium-mass nuclei, nucleon-nucleus optical potentials and elastic scattering cross sections have been computed with chiral forces
within the Self Consistent Green's Function (SCGF) approach \cite{scgf0,waldecker,jenkins,idini_prl} and the coupled-cluster method \cite{hagen2012c,gfccpap,gfccpap2}.

The optical potential plays an important role in reaction theory.
It is usual (and practical) in this context to reduce the many-body problem into a few-body one where only the most relevant degrees of freedom are retained\cite{ReactionsBook}. Correspondingly, the many-body Hamiltonian is replaced by a few-body Hamiltonian expressed in terms of optical potentials {\it i.e.} effective interactions between the particles considered at the few-body level.
Traditionally, optical potentials have been constructed by fitting to data, particularly data on $\beta$-stable isotopes \cite{KD,pheno}.
For instance, global phenomenological nucleon-nucleus potentials enable the description of
scattering processes for a large range of nuclei and projectile energies.
However, extrapolation of these phenomenological potentials to exotic regions of the nuclear chart are unreliable
and have uncontrolled uncertainties.
Moreover, since fitting to two-body elastic scattering data (as it is most often done) does not constrain the off-shell behavior of potentials\footnote{Two phase-equivalent potentials will reproduce the same elastic two-body scattering data
but may have different off-shell behavior.},
a dependence on the choice of potentials may arise in transfer reactions observables (and other reactions) as shown in  {\it e.g.} \cite{Titus_prc2014,Ross_prc2015,titus2}.
It is then critical, in order to advance the field of nuclear reactions and
notably for reactions with exotic nuclei
undertaken at rare-isotope-beam facilities\cite{frib_lab,fair}, to connect the 
optical potentials to an underlying microscopic theory of nuclei.
Since potentials derived from {\it ab initio} approaches are built up from fundamental nuclear interactions without tuning to data,
they may have a greater predictive power in regions of the nuclear chart that are unexplored experimentally.
Furthermore, they can guide new parametrization of phenomenological potentials by providing insights on form factors, energy-dependence
and dependence on the isospin-asymmetry of the target.

It is useful for pedagogical purpose and the introduction of key concepts, to start with the
derivation of the optical potential within the Feshbach projection formalism \cite{Feshbach:58,Feshbach:62}. Let us consider
the process of scattering of a nucleon on a target $A$. One can partition the  Hilbert space for this $A+1$ system  into  $\mathcal{P}$  the subspace
of elastic scattering states and $\mathcal{Q}$ the complementary subspace. Denoting $P$ and $Q$ the projectors
operators on respectively $\mathcal{P}$ and $\mathcal{Q}$, by construction one has $P+Q=Id$. We introduce $H$ the
Hamiltonian of the system and $E$ its energy. The optical potential describing the elastic scattering process
can be identified with the effective Hamiltonian $H^{eff}_{P}(E)$ acting in $P$, which by construction, reproduces
the eigenvalues of $H$ with a model wavefunction in $\mathcal{P}$. One can show that 
\begin{equation} \label{fesh}
 H^{eff}_{P}(E)=H_{PP}+H_{PQ}\frac{1}{E-H_{QQ}+i\eta}H_{QP}
\end {equation}
where  $H_{PP}\equiv PHP$, $H_{PQ}\equiv PHQ, \dots$ and $\eta \rightarrow 0^+$. The optical potential  $H^{eff}_{P}(E)$ is non-local
and from Eq.~\ref{fesh}, it is clear that it is also energy-dependent and complex. The imaginary (absorptive) component of the potential represents the loss of flux in the elastic channel due to the opening of other channels, for instance, the excitation of the target to a state of energy $E^{A}_{i}$ for $E>E^{A}_{i}$ or breakup channels. By adding the  Hilbert space of the $A-1$ system (hole states in the target)
in the formalism, it has been shown that the resulting optical potential corresponds
to the self-energy defined in Green’s function theory \cite{capuzzi}.
The particle part of the self-energy
is equivalent to the optical potential (\ref{fesh}), whereas the hole part describes the structure of the target.
By including information on  both the $(A+1)$- and $(A-1)$-system in the formalism, 
the Green's function approach, which will be used in this paper, provides a consistent treatment of scattering and structure.

In this article, we present some recent results for the {\it ab-initio} computation of  nucleon-nucleus optical potential for
medium-mass nuclei,  constructed by combining the Green’s function approach with the coupled-cluster method \cite{cc_review}. 
The coupled-cluster method is an efficient tool for the computation of ground- and low-lying excited states
in nuclei with a closed (sub-)shell structure and in their neighbors with $\pm 2$ nucleons.
By including complex continuum basis states in the formalism, it also provides a versatile framework to
consistently compute bound, resonant states and scattering processes \cite{hagen_cc_gam,hagen_oxy,CC_17F,cc_ni78,hagen2012c}.
In our approach, the optical potential is obtained by solving the Dyson equation after a direct computation
of the Green's function with the coupled-cluster method.
As we will see in Sec.~\ref{sec2}, the inclusion of complex continuum basis states enables also
a precise computation of Green's functions and optical  potentials.

We want to point out here that there has been a lot of work over the years to compute
optical potentials from various microscopic approaches. In the following, we mention some of the most
recent works dedicated to that goal (for a more exhaustive review we refer the reader to {\it e.g.} \cite{wim_review}).
The authors in \cite{blanchon1} have computed optical potentials for neutron and proton elastic scattering on $^{40}$Ca
based on the application of the self-consistent Hartree-Fock and Random-Phase Approximations to account for collective states in the target.
Using the phenomenological Gogny interaction,
a good reproduction of data for scattering at $\rm{E\leq 30}$ MeV has been reported in \cite{blanchon1}.
In \cite{holt1,holt2},  nucleon-nucleus  potentials are computed for finite nuclei from 
  a folding of optical potentials obtained by many-body perturbation theory calculations in nuclear matter with
 chiral forces. In these papers, several calcium isotopes are considered and an overall satisfactory agreement with  data is achieved.
For the scattering of nucleons at intermediate and high energy (E$\gtrsim$ 100 MeV) optical potentials can be derived
within the multiple scattering formalism \cite{amos,kerman_scat} where the optical potential is obtained based on the folding of the
nucleon-nucleon T-matrix or G-matrix with the nuclear density \cite{yamaguchi_bruckner,furumoto_g_matrix,vorabbi0}. Recent
applications of this approach, in which the nucleon-nucleus T-matrix and the density are computed consistently starting from the same chiral-EFT
interaction, have  been reported  and shown a successful reproduction of data \cite{gennari,burrows}.
In the Dispersive Optical Model \cite{wim_review,DOM,DOM2,DOM3}, a (semi-) phenomenological potential
is constructed by exploiting formal properties of the Green’s function, such as the dispersion relation,
which connects the real part and imaginary part of the potential \cite{mahaux_dom}. Applications of this data-driven approach have been made
using local and non-local form factors of the potential for Ca and Pb isotopes.

This paper is organized as follows. In Sec.~\ref{sec2}, we will briefly review the formalism to construct optical potentials
by combining the Green's function approach and the coupled-cluster method.
In Sec.~ \ref{sec_res}, recent results for neutron-$^{40,48}$Ca optical potentials
at negative and positive energies are presented. In Sec.~ \ref{challenge}, we will discuss  challenges and possible solutions
for the construction of fully predictive optical potentials with the coupled-cluster method. Finally, we will conclude in Sec.~\ref{conclusion}.
\section{Coupled Cluster Green's Function} \label{sec2}
In this part, we will briefly review the formalism for deriving  {\it ab-initio} nucleon-nucleus optical potentials
by combining the Green's function approach and the coupled-cluster method.
We start first by introducing below, key quantities of the Green's function formalism. 
\subsection{Green's Function and Dyson Equation}
Given a single-particle basis $\{|\alpha\rangle, |\beta\rangle,\ldots \}$, the  Green's function \cite{dickhoff} of a nucleus $A$
has matrix elements
\begin{eqnarray}
G(\alpha,\beta,E)&=&\langle \Psi_{0}|a_{\alpha}\frac{1}{E-(H-E^{A}_{gs})+i \eta} a^{\dagger}_{\beta}|\Psi_0\rangle + \langle \Psi_{0}|a^{\dagger}_{\beta}\frac{1}{E-(E^{A}_{gs}-H)-i \eta} a_{\alpha}|\Psi_0\rangle. 
\label{gf1}
\end{eqnarray}
Here, $H$ is the Hamiltonian and $|\Psi_0\rangle$ the ground state of $A$ with the energy $E^{A}_{gs}$ and by definition $\eta \rightarrow 0^+$.
 The operators $a^{\dagger}_{\alpha}$ and $a_{\beta}$ create and annihilate a fermion in the single-particle state
$\alpha$ and $\beta$, respectively. $\alpha$ is shorthand for the quantum numbers $\alpha = (n, l, j, j_z, \tau_{z})$ \footnote{
 $n, l, j, j_z, \tau_z$ label the radial quantum number, the orbital angular momentum, the total orbital momentum, its projection on the
   z-axis, and the isospin projection, respectively}. By inserting  completeness relations
 expressed with the eigenstates of the $A\pm 1$ systems in (\ref{gf1}), one  obtains the Lehmann representation of the Green’s function:
\begin{eqnarray}
G(\alpha,\beta,E)=
\sum_{i} \frac{\langle \Psi_{0}|a_{\alpha}|\Psi^{A+1}_{i}\rangle \langle \Psi^{A+1}_{i}|a^{\dagger}_{\beta}|\Psi_{0}\rangle}{E-(E^{A+1}_{i}-E^{A}_{gs})+i\eta} 
+\sum_j\frac{\langle \Psi_{0}|a^{\dagger}_{\beta}|\Psi^{A-1}_{j}\rangle \langle \Psi^{A-1}_j|a_{\alpha}|\Psi_{0}\rangle}{E-(E^{A}_{gs}-E^{A-1}_j)-i\eta}, 
\label{lehm}
\end{eqnarray}
where $|\Psi^{A+1}_i\rangle$ ($|\Psi^{A-1}_j\rangle$) is an eigenstate
of $H$ for the $A+1$ ($A-1$) system with energy $E^{A+1}_i$
($E^{A-1}_j$).  To simplify the notation, the completeness relations
are written in (\ref{lehm}) as discrete summations over the states in
the $A\pm 1$ systems. The Lehmann representation has the merit to reveal somewhat more clearly some of the information content
of the Green's Function. As one can see from (\ref{lehm}), the poles of the Green's function correspond to the energies of the eigenstates of $H$ in the $A\pm1$ systems. 

The Green's function fulfills the Dyson equation
\begin{eqnarray}
G(E)=G^{(0)}(E) +G^{0}(E)\Sigma^{*}(E)G(E) ,
\label{dys}
\end{eqnarray}
where $G^0(E)$ is the Green's function associated with a single-particle potential $U$ and  $\Sigma^{*}(E)$  the irreducible self energy.
The optical potential is given by
\begin{eqnarray}
V^{opt}(E)\equiv\Sigma^{*}(E)+U. \label{opt}
\end{eqnarray}
The potential $U$ is usually taken as the Hartree-Fock (HF) potential since the corresponding Green's function
is a first-order approximation to $G(E)$ in eq.~ (\ref{dys}). In our approach, since the Green's function
is directly computed with the coupled-cluster method and is input of Eq.~\ref{dys}, the resulting optical potential is independent of the choice
of $U$.

For $E^+\equiv E-E^{A}_{gs}\geq 0$, $V^{opt}(E)$ corresponds to the optical potential for the elastic scattering from the $A$-nucleon ground state\cite{dickhoff}. In other words, the scattering amplitude  $\xi_{E^+}({\bf r})=\langle \Psi_{0}|a_{{\bf r}}|\Psi_{E+}\rangle$
(here $|\Psi_{E^+}\rangle$ is the elastic scattering state of a nucleon on the target with the energy $E^+$ and
$a_{\bf{r}}$ is the annihilation operator of a particle at the position ${\bf{r}}$) fulfills the Schr\"odinger equation 
\begin{eqnarray}
-\frac{\hbar^2}{2\mu}\nabla^2\xi({\bf r})+\int d{ \bf r'} 
V^{opt}( {\bf r},{\bf r'},E) \xi({\bf r'})=E^+ \xi({\bf r}),
\label{schro}
\end{eqnarray}
where $\mu$ is the reduced mass of the nucleus-nucleon system. For simplicity, we have suppressed any spin and isospin labels.
The optical potential is non-local, energy-dependent and complex \cite{dickhoff} and for 
$E^+\geq 0$, its imaginary component describes the loss of flux in the elastic channels to other channels.
For $E^+ < 0$, Eq.~\ref{schro} admits a discrete number of physical solution at $E_n=E_n^{A+1}-E^{A}_{gs}$, which corresponds
to the bound states energies in A+1. In that case, the solutions are given by the overlap
$\xi_{n}({\bf r})=\langle \Psi_{0}|a_{\bf{r}}|\Psi^{A+1}_{n}\rangle$ where $|\Psi^{A+1}_{n}\rangle$ is a bound state
of energy $E_n^{A+1}$ in the $A+1$ system
\footnote{Similarly, for $E=E^{A}_{gs}-E_{n}^{A-1}$, the solution of the optical potential $V^{opt}(E)$
  are the radial overlap $\xi^{-}_{n}({\bf r})=\langle \Psi_{0}|a^{\dagger}_{\bf{r}}|\Psi^{A-1}_{n}\rangle$  \cite{dickhoff}.}.

In the following section, we present the main steps involved in the computation of the Green's function with the
coupled-cluster method.
\subsection{Coupled-cluster Green's Function} \label{cc_section}
We start with the computation  of the ground state $|\Psi_0\rangle$ of the $A$-nucleon system.
Working in the laboratory frame, the intrinsic Hamiltonian  reads
\begin{eqnarray}
H=\sum^{A}_{i=1}\frac{\vec{p_i}^2}{2m}
-\frac{\vec{P}^2}{2mA} +\sum_{i<j} V_{ij} +\sum_{i<j<k} V_{ijk}, 
\label {hami}
\end{eqnarray}
with $\vec{p_i}$ the momentum of nucleon $i$ of mass $m$  and $\vec{P}=\sum_{i=1}^{A}\vec{p_i}$
 the momentum associated with the center of
 mass motion. The terms $V_{ij}$ and $V_{ijk}$ are nucleon-nucleon $(NN)$ and three-nucleon forces $(3NFs)$, respectively.
  It is useful to rewrite the Hamiltonian as
\begin{eqnarray}
H=\sum^{A}_{i=1}\frac{\vec{p}_i^2}{2m} \left (1-\frac{1}{A}\right)   +\sum_{i<j} \left ( V_{ij} -\frac{\vec{p}_i\vec {p}_j}{mA} \right )  +\sum_{i<j<k}V_{ijk}  ,~~~~ \label {hami2}
\end{eqnarray}
where one separates the one-body and two- (three-)body contributions.
The single-particle basis solution of the HF potential generated by $H$ in eq.~(\ref{hami2})
is a good starting point for coupled-cluster calculations.
Denoting by $|\Phi_{0}\rangle$ the HF state, the ground state of the target
is represented as
\begin{eqnarray}
|\Psi_{0}\rangle=e^T|\Phi_{0}\rangle \label{cc1} , 
\end{eqnarray}
where $T$ denotes the cluster operator
\begin{eqnarray}
T &=& T_1+T_2+\dots= \sum_{i,a}t_i^a a^{\dagger}_a a_i+\frac{1}{4}\sum_{ijab}t_{ij}^{ab}t_{ijab}a^{\dagger}_aa^{\dagger}_ba_ja_i +\ldots .
\label{t_cluster}
\end{eqnarray}
The operators $T_1$ and $T_2$ induce $1p-1h$ and $2p-2h$ excitations of the reference state, respectively.
Here, the single-particle states $i, j, ...$ refer to hole
states occupied in the reference state $|\Phi_0\rangle$  while $a, b, ...$
denote valence states above the reference state. In practice, the expansion (\ref{t_cluster}) is truncated. In the coupled
cluster method with singles and doubles (CCSD) all operators $T_i$
with $i >$ 2 are neglected.  In that case, the ground-state energy and the amplitudes
$t_i^a, t_{ij}^{ab}$ are obtained by projecting the state (\ref{cc1}) on the
reference state and on all $1p$-$1h$ and $2p$-$2h$ configurations for which
\begin{eqnarray}
  \label{ccsd}
  \nonumber
  \langle \Phi_0|\overline{H}|\Phi_0\rangle&=&E ,  \\
  \nonumber
  \langle \Phi_i^a|\overline{H}|\Phi_0\rangle&=&0 ,  \\
 \langle \Phi_{ij}^{ab}|\overline{H}|\Phi_0\rangle&=&0 .
\end{eqnarray}
Here,
\begin{eqnarray}
  \overline{H}&\equiv& e^{-T}He^T = H + \left[H,T\right] +{1\over 2!}\left[\left[H,T\right],T\right] + \ldots \label{bch}
\end{eqnarray}
denotes the similarity transformed Hamiltonian, which is computed by
making use of the Baker-Campbell-Hausdorff expansion \cite{cc_review}.
For two-body forces and in the CCSD approximation, this expansion  terminates at fourfold nested
commutators \footnote{The $3NFs$ component $V_{ijk}$ of the Hamiltonian  in (\ref {hami2}) is truncated  at the normal-ordered two-body level in the HF basis (see Sec.~\ref{sec_res}).}. The CCSD equations (\ref{ccsd}) show that the CCSD ground state is an eigenstate
of the similarity-transformed Hamiltonian $\bar{H}=e^{-T}He^{T}$ in the space of $0p-0h$, $1p-1h$, $2p-2h$ configurations.
The operator $e^T$ being not unitary, $\bar{H}$ is not Hermitian.
As a consequence, its left- and right-eigenvectors form a bi-orthonormal set \cite{cc_review}.

Denoting $\langle \Phi_{0,L}|$ the left eigenvector for the ground state of $A$, we can now write the matrix elements of the
coupled cluster Green's function $G^{cc}$ as
\begin{eqnarray}
G^{CC}(\alpha,\beta,E) \equiv \langle \Phi_{0,L}|\overline{a_{\alpha}}\frac{1}{E-(\overline{H}-E^{A}_{gs})+i\eta}\overline{a^{\dagger}_{\beta}}|\Phi_{0}\rangle +\langle \Phi_{0,L}|\overline{a^{\dagger}_{\beta}}\frac{1}{E-(E^{A}_{gs}-\overline{H})-i\eta}\overline{a_{\alpha}}|\Phi_{0}\rangle .
\label{gfcc}
\end{eqnarray}
Here, $\overline{a_{\alpha}}=e^{-T}a_{\alpha}e^T$ and
$\overline{a^{\dagger}_{\beta}}=e^{-T}a^{\dagger}_{\beta}e^T$ are the
similarity-transformed annihilation and creation operators,
respectively. These are computed with the  Baker-Campbell-Hausdorff expansion (\ref{bch}).

In principle, the Green’s function could be computed from the Lehman decomposition (\ref{lehm}) with the solutions of the particle-attached equation of-motion (PA-EOM) and particle-removed equation-of motion (PR-EOM) for the  $A+1$ and $A-1$ sytems, respectively \cite{cc_review}. However, as the sum over all states  in Eq.~(\ref{lehm}) involves also eigenstates in the
continuum,  this approach is difficult to pursue in practice. Instead, we make use of the 
 Lanczos continued fraction technique, which allows for an efficient and numerically stable computation of the Green's function \cite{gfccpap,lanc_method,lanc_method2,dagotto1994,hallberg1995,haxton2005}.

By definition of the Green's function, the parameter $\eta$ in the matrix elements (\ref{gf1}) is such that $\eta \rightarrow 0^+$.
However, in  this limit, because of the appearance
of poles at energies $E=(E^{A+1}_i-E^{A}_{gs})$ in the Green's function (see Eq.~(\ref{lehm})), the  calculation of  optical potential for elastic scattering becomes numerically unstable.
In order to resolve this issue, we compute an analytic continuation of the Green's function in the complex-energy plane
by working in a Berggren basis \cite{berggren1968,michel2002,idbetan2002,hagen2004,hagen2006,dmrg,nick,fossez18}
(generated by the HF potential) that includes bound, resonant, and complex-continuum
states. The  solutions of the (PA-EOM) and (PR-EOM) in the Berggren basis, {\it i.e} the eigenstates of the $A\pm 1$ systems,
are either bound, resonant or complex-scattering states. In other words, the poles of the analytically
continued Green's function are located either at negative real or complex
energy. As a result, the Green's function matrix elements for $E\geq0$
smoothly converge to a finite value as $\eta \rightarrow 0^+$ (this is illustrated below in Fig.~\ref{fig_dens}).

The scattering states entering the Berggren basis are defined along a contour
$L^+$ in the fourth quadrant of the complex momentum plane,
below the resonant single-particle states. According to the Cauchy
theorem, the shape of the contour $L^+$ is not important,
under the condition that  all resonant states lie between the contour and the real
momentum axis.  The Berggren completeness reads
\begin{eqnarray}
\sum_{i}|u_i\rangle\langle \tilde{u_i}|+\int_{L^{+}}dk|u(k)\rangle\langle \tilde{u(k)}|= \hat{{1}}, 
\end{eqnarray}
where $|u_i\rangle$ are discrete states corresponding to bound and
resonant solutions of the single-particle potential, and
$|u(k)\rangle$ are complex-energy scattering states along the
complex-contour $L^+$. In practice, the integral along the complex
continuum is discretized yielding a finite discrete basis set.

\begin{figure}[h!]
\begin{center}
\includegraphics[width=10cm]{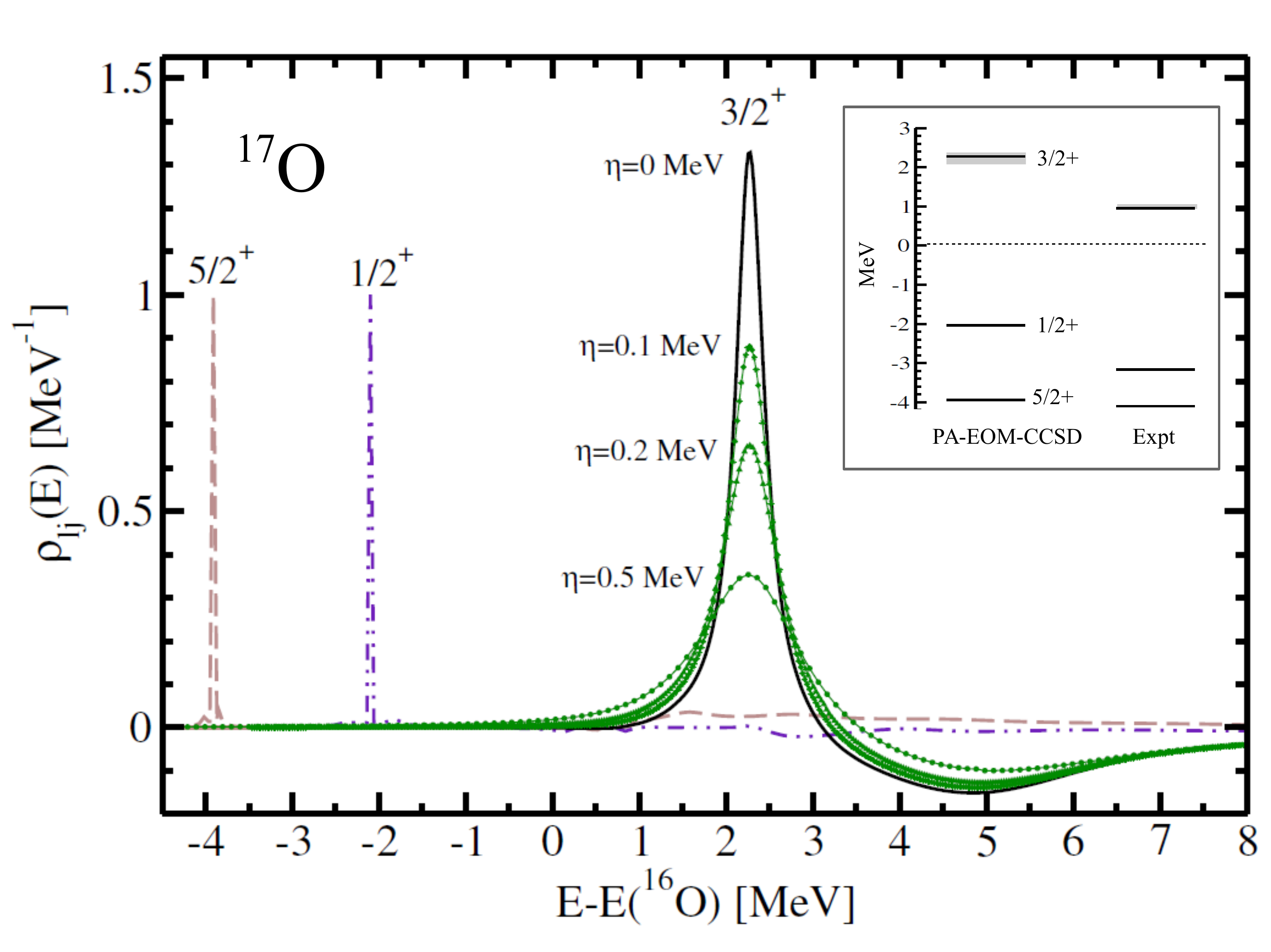}
\end{center}
\caption{Computed level densities in $^{17}$O. For the $J^{\pi}=3/2^+$ level density, results are  shown for several values
  of the parameter $\eta$ to illustrate the smooth convergence pattern for $\eta\rightarrow 0$. 
  The inset shows the energies of the ground state, first excited and $3/2^{+}$ resonant states 
  in  $^{17}$O calculated at the PA-EOM-CCSD truncation level (see text for details).}
\label{fig_dens}
\end{figure}

In Fig.~\ref{fig_dens}, we illustrate the numerical stability  provided by
the use of the  Berggren basis for the computation of the Green's function. 
We are interested in the level density \cite{shlomo_dens,vig}  
\begin{eqnarray}\label{dens}
\rho_{lj}(E)=-\frac{1}{\pi}Tr \left [Im (G_{lj}(E)-G_{lj}^{(0)}(E)) \right],
\end{eqnarray}
where $G_{lj}(E)$ and $G^{0}_{lj}(E)$ are respectively the component of the Green's functions and the HF Green's function
in the $(l,j)$ partial wave \footnote{Since the Green's functions are here defined by adding (and removing) a
nucleon from  the $0^+$ ground state in the target $A$, the quantum number $(l,j)$ are conserved.}. 
We show in Fig.~\ref{fig_dens}, the $J^{\pi}=3/2^{+}$ level density in $^{17}$O calculated with the $\rm{NNLO_{sat}}$ interaction.
The ground state in $^{16}$O is computed at the CCSD level while the Green's function is computed with the PA-EOM
and PR-EOM Lanczos vectors truncated at the $2p-1h$ and $1p-2h$ excitation level, respectively (other details of the calculation are also the same as in Sec.~\ref{sec_res}). As $\eta$ approaches 0, the level density smoothly converges, and the position of the peak at $\eta=0$ corresponds, as expected, to
the position of the  $J^{\pi}=3/2^{+}$ resonance in $^{17}$O (see inset in Fig.~\ref{fig_dens}, which shows the PA-EOM-CCSD energies in $^{17}$O).
For completeness, we also show the  $J^{\pi}=5/2^{+},1/2^{+}$ level densities. In these cases, the level density at negative energies
are  equal to a Dirac delta function  peaked at respectively the ground state and first excited state energies
in $^{17}$O (see inset in Fig.~\ref{fig_dens}). For purpose of illustration in  Fig.~\ref{fig_dens}, we
have used a finite value of $\eta$ for the  $J^{\pi}=5/2^{+},1/2^{+}$ densities and set the height of the corresponding peaks to 1. 
\section{Selected results}\label{sec_res}
We now show in this section a few results of the computation of neutron optical potentials
for the double-magic nuclei $\rm{^{40}Ca}$ and  $\rm{^{48}Ca}$.

All calculations presented here are performed using the $\rm{NNLO_{sat}}$ chiral interaction \cite{n2losat},
which reproduces the binding energy and charge radius of both systems \cite{hagen2015,garciaruiz2016}.
We want to point out here that a proper reproduction of the distribution of nuclear
matter, and, more specifically, nuclear radii is critical in order to obtain an accurate account of reactions observables. All results are
obtained from coupled-cluster calculations truncated at the CCSD level, while the Lanczos vectors in the PA-EOM
 (PR-EOM) have been truncated at the $2p-1h$ ($1p-2h$) excitation level.
Since the computation of the Green's function is performed using the laboratory coordinates (the Hamiltonian $H$ in Eq.~(\ref{hami2}) is defined with these coordinates), the calculated optical potential is identified 
with the optical potential in the relative coordinates of the $\rm{n-^{A}Ca}$ system. This identification will result in a small 
error, which is a decreasing function of the target mass number $A$  \cite{gfccpap,gfccpap2} (see also Sec.~\ref{challenge}).

The HF calculations are performed in a mixed basis of harmonic oscillator
and Berggren states, depending on the partial wave. 
The $\rm{NNLO_{sat}}$ interaction contains  two-body and three-body terms. Denoting $N_2$ and $N_3$
the cutoffs in the harmonic oscillator (HO) basis of respectively, the two-body and  three-body part of the interaction,
we set $N_2=N_3=N_{max}$ except for the most extensive calculations where $N_2 = 14$ and $N_3 = 16$.
Finally, we truncate the three-nucleon forces at the normal-ordered two-body level in the HF basis. This approximation has been shown to work well in light- and medium mass nuclei \cite{hagen2007a,roth2012}. The
harmonic oscillator frequency is kept fixed at $\hbar \omega=$16 MeV (for more details see \cite{gfccpap,gfccpap2}).
\begin{figure}[h!]
\begin{center}
\includegraphics[width=11cm]{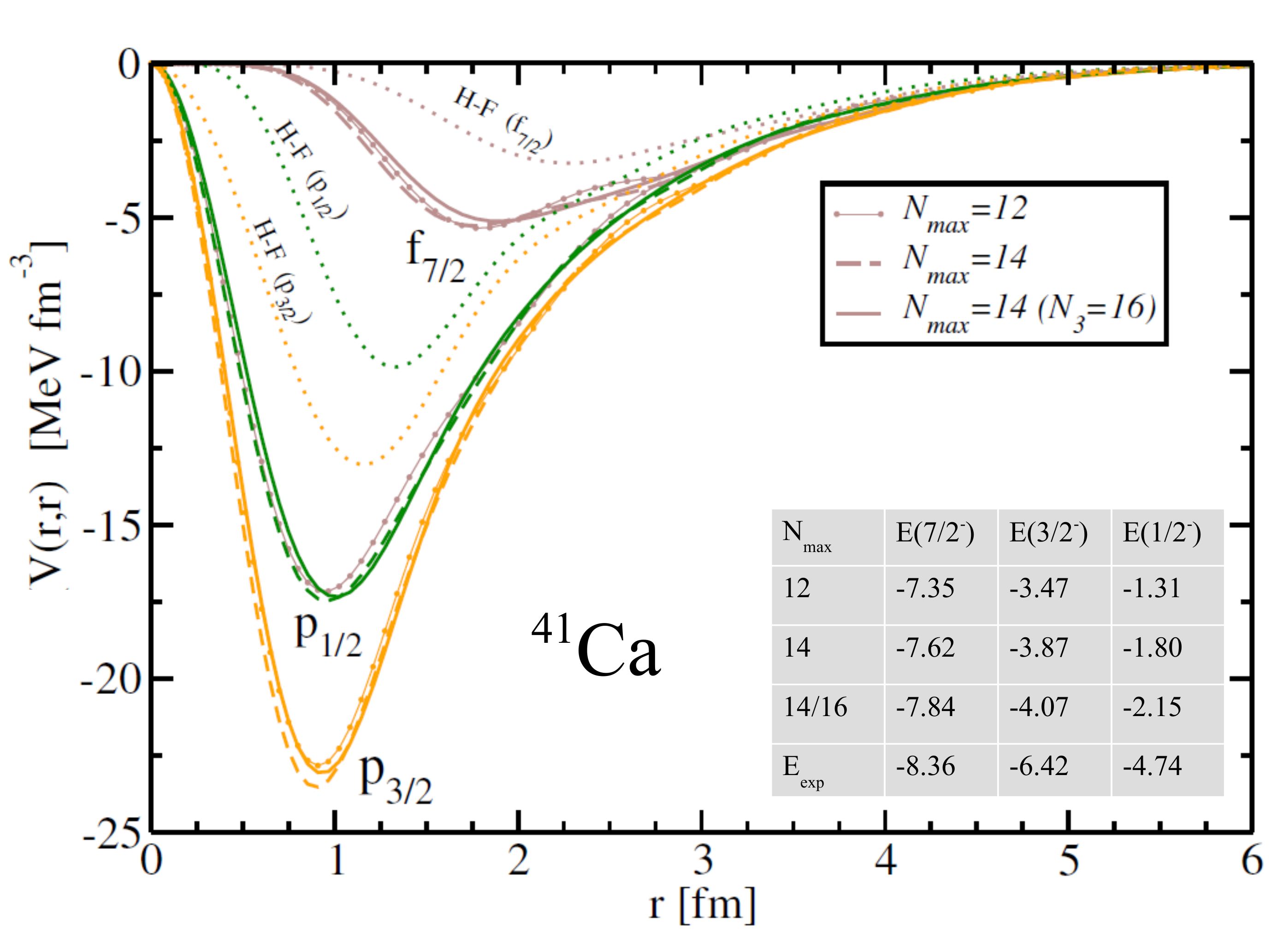}
\end{center}
\caption{Diagonal part of the $\rm{n+^{40}Ca}$ optical potential for the bound states in $\rm{^{41}Ca}$ computed
  with the $\rm{NNLO_{sat}}$ interaction. Results are shown  for several values of $N_{max}$ and the corresponding bound state energies
  (with respect to the $^{40}$Ca ground state)
  are shown in the table (in MeV). The components of the HF potential in the associated partial waves are shown 
 for  $(N_{max},N_3)=(14,16)$ (see text for details).}\label{fig_calcium_pot}
\end{figure}

We start with the computation of the $\rm{n+^{40}Ca}$ optical potentials associated with the bound states in $^{41}$Ca.
At the PA-EOM-CCSD level of truncation considered here, there are only three bound states supported by the $\rm{NNLO_{sat}}$ Hamiltonian.
In order to show the convergence pattern of the potentials, we present in Fig.~(\ref{fig_calcium_pot}) results at several values of $N_{max}$
with the corresponding  bound state energies. We present the diagonal part of the potentials, and for comparison
the HF potential (for $(N_{max} , N_3 )=(14,16)$) in each partial wave is also shown in Fig.~(\ref{fig_calcium_pot}).
 The energies are  shown in the table in Fig.~(\ref{fig_calcium_pot}) along with the experimental values.  As expected, the convergence
of  energies is slower for  higher-energy states. The difference between the $\rm{^{41}Ca}$ energies at
$(N_{max},N_3)$=(14,14) and (14,16) is $\sim$ 220 keV in the case of the ground-state, whereas it is $\sim$ 350 keV in the case
of the $J^{\pi} = 1/2^-$ second excited state. Even though the absolute binding energy
is underestimated in the CCSD approximation, when compared to experiment (the CCSD binding energy of $^{40}$Ca is 299.28 MeV for ($N_{max},N_3$ ) = (14, 16), whereas the experimental value is 342.05 MeV), the neutron separation energies
are consistently within 600 keV of the experimental values for $^{40,48}$Ca \footnote{By including both perturbative triple excitations and perturbative estimates for the neglected residual 3NFs (3NF terms beyond the normal-ordered
two-body approximation), a good agreement with experimental binding energies can be obtained for $^{40,48}$Ca\cite{hagen2015}}.
The eigenenergies of these potentials are equal, by construction,
to the bound states energies when using the effective mass $mA/(A-1)$ instead of the actual reduced mass.
This can be traced to Eq. (\ref{hami2}) where the effective mass associated with the one-body kinetic operator is
equal to $mA/(A-1)$ (see also Sec.~\ref{challenge}). 

%

\begin{figure}[h!]
\begin{center}
\includegraphics[width=12cm]{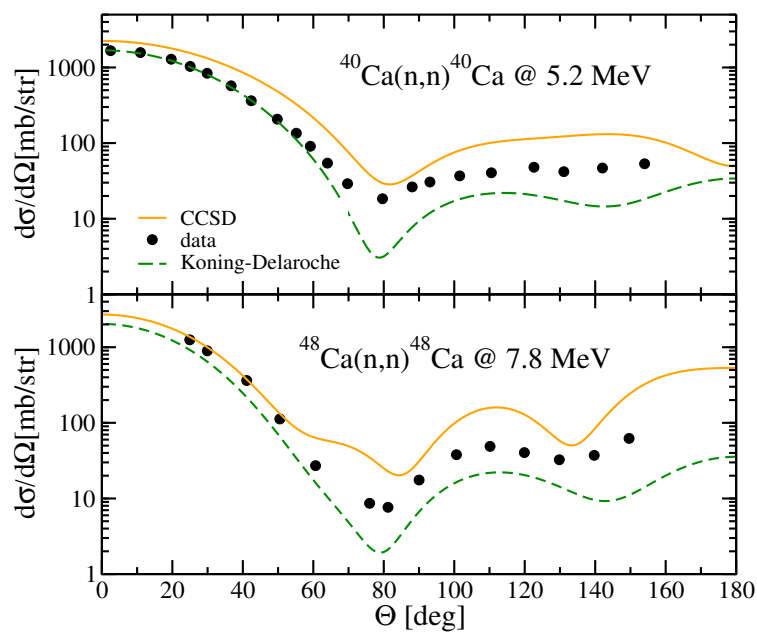}
\end{center}
\caption{Differential elastic cross section for  $\rm{^{40}Ca(n,n) ^{40} Ca} $ at 5.2 MeV (top)  and  $\rm{^{48}Ca(n,n) ^{48} Ca} $ at 7.8 MeV (bottom) calculated with the  $\rm{NNLO_{sat}}$ interaction. Results obtained with the phenomenological Koning-Delaroche potential
potential are shown (dashed line) for comparison. Data points are taken from \cite{KD} (errors
on the data are smaller than the symbols).}\label{fig_calcium}
\end{figure}

We now consider the neutron elastic scattering on $\rm{^{40}Ca}$ and  $\rm{^{48}Ca}$.
The phase shift is computed in each partial wave with the optical potential calculated in the largest space  $(N_{max},N_3)=(14,16)$.
The angular distributions are then obtained by summing the contributions from each partial wave.
Figure ~\ref{fig_calcium} shows the resulting differential elastic cross section
for $\rm{^{40}Ca(n,n)^{40}Ca}$ at 5.2 MeV and  $\rm{^{48}Ca(n,n)^{48}Ca}$ at 7.8
MeV. We find that at these energies the inclusion of partial waves with angular momentum $L\leq 5$ and
$L \leq 6$ is sufficient for $^{40}$Ca and $^{48}$Ca respectively, the
contribution of partial waves with higher $L$ being negligible (see also the computations of elastic scattering on
$^{40,48}$Ca at other energies  in \cite{gfccpap2}).
The angular distributions obtained with the phenomenological Koning Delaroche (KD) potential \cite{KD} and the measured cross sections
are also shown in  Fig.~\ref{fig_calcium} for comparison. As Fig.~\ref{fig_calcium} indicates, the data at small angle
where the cross section is larger, are well reproduced
for $\rm{^{48}Ca}$ whereas the computed cross section is slightly above the data for $\rm{^{40}Ca}$.
Overall, the shape of the experimental cross sections and the positions of the minima are well reproduced for both nuclei,
as expected from the correct reproduction of matter densities in $\rm{^{40,48}}$Ca by the $\rm{NNLO_{sat}}$ interaction.

The experimental energy of the first two excited-states  in $^{40}$Ca, namely  E$(0^{+})$=3.35 MeV and  E$(3^-)$=3.74 MeV
are below the scattering energy $\rm{E_{scat}}$=5.2 MeV of the elastic process $\rm{^{40}Ca(n,n)^{40}Ca}$ shown in Fig.~\ref{fig_calcium}.
In other words, the channels for excitation of the $^{40}$Ca target  are open at this scattering energy. This should result in a loss of flux
in the initial elastic channel and the corresponding occurrence of an absorptive imaginary part in the phase shifts.
The first excited $0^+$ state, which has a strong $4p-4h$ components, cannot be properly reproduced at the truncation level 
considered here: its computed energy, solution of the EOM-CCSD equations, is $\sim$16 MeV above the ground state.
On the other hand, the $3^-$ excited state is well reproduced with $E_{EOM-CCSD} (3^{-})$ =3.94 MeV.
Nevertheless, we have found that the computed absorption is practically negligible and none of the computed phase shifts at
$\rm{E_{scat}}$=5.2 MeV have a significant imaginary part.
A similar pattern happens for $\rm{^{48}Ca(n,n)^{48}Ca}$ at 7.8 MeV: in that case, the first
excited state E($2^+$)=3.83 MeV is fairly well reproduced, the computed value is $E_{EOM-CCSD}(2^+)$ =4.65 MeV,
but again the  absorption in that case is negligible too.

Although some excited states below the scattering energy are reproduced by the EOM-CCSD calculations, the
absorption is negligible in both situations. This suggests that at the level of truncation considered here, namely $2p-1h$ above the CCSD ground state, the computed wavefunctions are not correlated enough (in the perturbative expansion of the Dyson equation Eq.~ (\ref{dys}),
the absorption appears at second-order, beyond the HF contribution \cite{dickhoff}). In other words, at these energies, the computed level density (\ref{dens}) in the $\rm{n+^{A}Ca}$ system
is too small. We have observed that only at higher energy E$\gtrsim$20 MeV the absorption starts to increase significantly 
(a similar pattern can be seen in Fig.~4 of \cite{gfccpap} for the CCSD computation of $\rm{n+^{16}O}$ optical potential).
It is possible to increase artificially the absorption by using a finite value of $\eta$ in Eq.~(\ref{gfcc}).
This amounts to increasing the correlations content of the coupled-cluster wavefunctions and 
as shown in \cite{gfccpap,gfccpap2}, the computed elastic cross section in that case will decrease.
In Sec.\ref{challenge},  we will return to this lack of absorption in the computed potential. 

We should  emphasize here that the computation of the optical potential with the coupled-cluster method is carried out without any free parameter. It is then not
surprising that it does not allow for the same quality of reproduction of data as a phenomenological potential such as the KD interaction (see Fig.~\ref{fig_calcium}). But still, since microscopic optical potentials are built up from fundamental nuclear interactions without tuning to data, they may yield guidance for parameterizations of phenomenological potential, by providing information on the form factor, energy dependence
and dependence on the isospin asymmetry of the targets. A recent series of studies has shown that non-locality can affect transfer reaction observables (e.g. \cite{Titus_prc2014,Ross_prc2015,titus2})
and it is expected that it can equally affect other reaction channels. Microscopic potential
can provide guidance on this aspect of the optical potential.
Keeping in mind that a potential is not an observable and is not uniquely defined (for a given potential, it is possible to modify its high-energy component with a unitary transformation without affecting experimental predictions \cite{bogner2010,arellano}), we focus
in the following  on the non-locality of the CCSD optical potential. 
\begin{figure}[h!]
\begin{center}
\includegraphics[width=11cm]{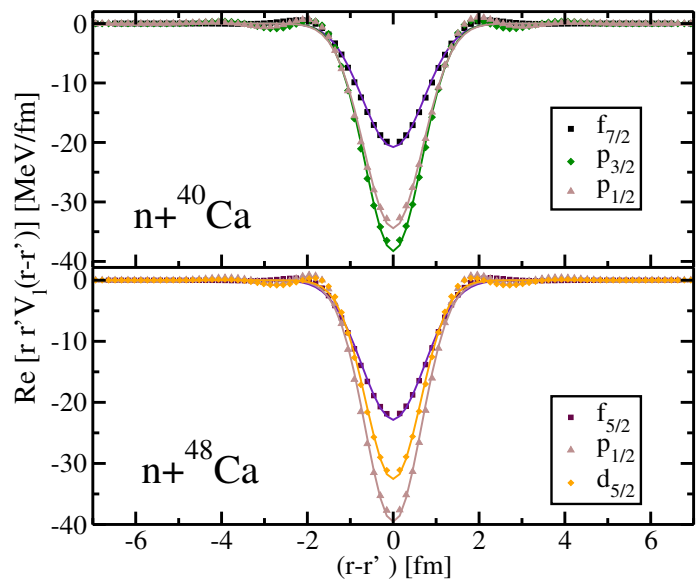}
\end{center}
\caption{Real part of the neutron potential in several partial waves for $^{40,48}$Ca at respectively 5.2 and 7.8 MeV.
  The potentials are shown at fixed values of $R$ (equal to the charge radius in both nuclei) and  as a function of $r-r'$.
  Symbols corresponds to the calculated potentials and the lines are the results of a fit with a Gaussian form factor (see text for details).}
\label{fig_non_loc}
\end{figure}

 We plot in Fig.~\ref{fig_non_loc}, the $\rm{n+^{40,48}Ca}$ potentials in several partial waves, at a fixed value of $R=(r+r')/2$
 and as a function of $r-r'$. We fix $R$ to be equal to the charge radius in both nuclei, namely 3.48 fm and 3.46 fm for respectively
 $^{40}$Ca and $^{48}$Ca \cite{n2losat}. We consider the same energy as previously, namely 5.2 MeV for $^{40}$Ca and 7.8 MeV for $^{48}$Ca.
 A fit of the potential using a Gaussian form factor, is also shown in Fig.~\ref{fig_non_loc}.
 As one can see,  the shape of potentials in Fig.~\ref{fig_non_loc} are well reproduced by the fit.
 For $^{40}$Ca, the values of the range $\beta$ of the fitted Gaussian somehow varies slightly  with the partial wave:
  we obtain $\beta=1.02, 0.94, 0.98$ fm for the $f_{7/2}, p_{3/2}$ and
 $p_{1/2}$ component of the potentials, respectively. For $^{48}$Ca, $\beta=1.04, 0.93, 0.91$ fm for  the $f_{5/2}$, $p_{1/2}$ and
 and $d_{5/2}$ partial waves, respectively. We have observed even smaller variations of the range with the energies although
a more exhaustive study would be required to draw definitive conclusion about the dependence of $\beta$
on the value of $R$ and the energy. Nevertheless, in all cases, the non-local pattern of the optical potential display a Gaussian
dependence, which corresponds to the choice made for the non-local form factor 
in the phenomenological potentials by Perey and Buck \cite{Perey:62}.
Note that due to the non-hermiticity of the Coupled Cluster Hamiltonian (see Sec.~\ref{cc_section}) the
potential is slightly non symmetric in $r$ and $r'$. However since this effect is small \cite{gfccpap,gfccpap2}, it  is hardly noticeable in Fig.~\ref{fig_non_loc}.
\section{Challenges} \label{challenge}
In this section, we discuss some challenges 
and  possible solutions for the development of fully predictive ab-initio optical potentials with the coupled-cluster method.

We saw in the previous section that with the {\it ab-initio} optical potentials computed at the CCSD level,
one can arrive at an overall fair reproduction of data for medium-mass
nuclei. However, the absorptive part of the potential was shown to be
negligible at low energy. This lack of absorption was
linked to neglected configurations  in the computed  Green's function.

Currently, {\it ab-initio} computation of optical potentials for medium-mass nuclei using chiral $NN$ and 3NFs, 
 have only been performed with the coupled-cluster method and the Self Consistent Green's Function (SCGF) method \cite{idini_prl}.
The SCGF is based on an iterative solution of the Dyson equation performed until a self-consistency between the input Green's function
and the result of the Dyson equation has been reached \cite{scgf0}. In \cite{idini_prl}, the authors compute neutron optical potential
for $^{16}$O and $^{40}$Ca with the $\rm{NNLO_{sat}}$ interaction and include up to $2p-1h$ configurations in the Green's function.
In that work, the minima in the elastic cross sections are well reproduced for both systems, and as in the CCSD computation of the potential, an
overall lack of absorption was observed and attributed to neglected configurations in the model space.

The natural next step to address the lack of absorption at the CCSD level  would be to include higher-order correlations
in the Green's function by considering  next order excitations in the coupled-cluster calculations, namely triple corrections.
One should expect in that case an increased level density in the $A+1$ system and as a result, a larger absorptive part of the
optical potential.  Coupled-cluster calculations with triple corrections are routinely used for nuclear spectroscopy \cite{cc_review} 
and have recently been implemented in the computation of the dipole polarizability of $^{48}$Ca \cite{miorelli2018}.
In that paper, the authors show that by including $3p-3h$ excitations in the computation of the nuclear response function to an
electromagnetic probe (the Green's function is a similar object since it is the response function to the addition/removal of a nucleon), the
results improve over previous computations  at CCSD.

For most nuclei, and particularly for heavier systems, there are many compound-nucleus
resonances above the particle threshold. Since these states consist of a high number of particle-hole excitations they
cannot be reproduced accurately by {\it ab-initio} methods and are usually best described by a stochastic approach \cite{mitchell2010}.
In order to account for the formation of the compound nucleus and the resulting loss of flux in the elastic channel,
one could  add a polarization term to the {\it ab-initio} potential. A possible way to compute this
term would be to use Random Matrix Theory to generate an effective Hamiltonian belonging to a Gaussian Orthogonal Ensemble \cite{fanto}.

Since the coupled-cluster Green's function is computed in the laboratory frame,
the optical potential solution of the Dyson equation is defined with respect to the origin of that frame $O$. As mentioned in Sec.~\ref{sec_res},  we have identified this potential with the potential in the relative $n-A$ coordinate.
For the medium-mass nuclei considered here, this prescription creates a small error, which decreases with $A$
\cite{gfccpap,gfccpap2,ron}.
For light systems,  a correction to the optical potential becomes necessary to account for the identification between laboratory and
relative coordinates.
It has been demonstrated that the coupled cluster wavefunction factorizes to a very good approximation  into a product of an
intrinsic wave function and a Gaussian in the center-of-mass coordinate \cite{hagen2009a}.
Since both the potential and the center-of-mass wavefunction of the target are computed in the laboratory frame, it seems reasonable to suggest that such a correction could be introduced in the form of a  folding of the potential with the center-of-mass wavefunction
 (nevertheless, such a prescription would have to be worked out and checked).
Another possible way to introduce a correction of the potential could be to use the integral method utilized in the  GFMC approach ( see {\it e.g} \cite{nollett_int}) for  computation of overlap functions (see also {\it e.g.} \cite{pink_satch,timo}).
 \section{Summary}\label{conclusion}
 In this article, we have presented  recent developments in the computation of nucleon-nucleus optical potential
 constructed by combining the Green's function  and  the coupled-cluster method. 
  A key element in this approach is the use of the Berggren basis,  which enables a consistent description of bound, resonant states and scattering process of the  (nucleon-target) system and at the same  time, allows to properly deal with the poles of the Green’s function on the real energy axis.
 
 We have shown results for optical potentials at negative and positive energy for the double magic systems $^{40}$Ca and $^{48}$Ca
 using a chiral $NN$ and 3NFs that reproduces the binding energy and charge radii in  both systems.
 We pointed out that a proper reproduction of the distribution of nuclear
 matter, and, more specifically, nuclear radii, by the Hamiltonian, is essential to give an accurate account of reaction observables.
 At the truncation level considered here, namely  $2p-2h$ and $2p-1h$ / $2h-1p$ in the computation of
 the target and the Green's function, respectively,  an overall fair agreement with data was obtained.
 Nevertheless, in that case,  the optical potential at positive energy
 suffers from a lack of absorption, which stems from the neglect of higher-order configurations.
 In (near) future development, higher-order excitations in the coupled-cluster  expansion will be included to address this issue.

 In the future, the Green's function formalism and coupled-cluster method could be combined
 for applications to other reaction channels such as transfer, capture, breakup and charge-exchange.
 Another possible approach toward the {\it ab-initio} computation  of transfer reactions with medium-mass nuclei
 is the Green's Function Transfer (GFT) method \cite{dp2}.  Using the optical potential and Green's function computed with the coupled-cluster method as input of the GFT equations, as well as phenomenological ingredients,
 a very good reproduction of data for populating the ground states in $^{41,49}$Ca
 was obtained with this approach.
Although the current implementations of the GFT method require phenomenological inputs, future extensions of the
formalism should allow {\it ab-initio} computation of transfer reactions \cite{dp2}.
\section*{Acknowledgments}
The author would like to thank his collaborators  P. Danielewicz, G. Hagen, G. R. Jansen, F. M. Nunes, and T. Papenbrock for their contributions to the studies presented in this work.
\bibliographystyle{frontiersinHLTHFPHY} 
\bibliography{refs}
\end{document}